# What is Human-Centeredness in Human-Centered AI? Development of Human-Centeredness Framework and AI Practitioners' Perspectives

**Aung Pyae**

International School of Engineering, Faculty of Engineering, Chulalongkorn University, Bangkok, Thailand
aung.p@chula.ac.th

There is no consensus on what constitutes human-centeredness in AI, and existing frameworks lack empirical validation. This study addresses this gap by developing a hierarchical framework of 26 attributes of human-centeredness, validated through practitioner input. The framework prioritizes ethical foundations (e.g., fairness, transparency), usability, and emotional intelligence, organized into four tiers: ethical foundations, usability, emotional and cognitive dimensions, and personalization. By integrating theoretical insights with empirical data, this work offers actionable guidance for AI practitioners, promoting inclusive design, rigorous ethical standards, and iterative user feedback. The framework provides a robust foundation for creating AI systems that enhance human well-being and align with societal values. Future research should explore how these attributes evolve across cultural and industrial contexts, ensuring the framework remains relevant as AI technologies advance.

## 1 INTRODUCTION

Recent advancements in artificial intelligence (AI) have led to the widespread integration of AI-driven applications into everyday life, necessitating a shift from purely technical considerations to a more holistic approach that accounts for human experiences. In response to this imperative, Human-Centered AI (HCAI) has emerged as a paradigm that prioritizes human well-being, aligns AI technologies with human values, and enhances user experience [Régis et al. 2024]. Drawing from interdisciplinary fields such as AI, Human-Computer Interaction (HCI), Human-Centered Design (HCD), and User Experience Design (UxD), HCAI bridges the domains of computer science, psychology, and engineering. At its core, HCAI advocates for collaborative, ethical, and human-centric design practices, ensuring that AI systems are not only technologically robust but also socially responsible.

The conceptual foundations of HCAI can be traced back to early HCI theories that emphasized human-technology collaboration. Pioneering work by [Engelbart 1962] and [Licklider 1960] laid the groundwork for augmenting human capabilities through technology, establishing foundational principles of user-centric design [Card et al. 1986]. Even during the AI winter (1970s–1990s), the field of HCI continued to evolve, with significant contributions such as Shneiderman's direct manipulation interfaces [Shneiderman 1983] and Norman's cognitive models [Norman 1986]. The formalization of iterative design processes centered on user needs was further advanced by IDEO's Human-Centered Design Toolkit [IDEO 2011]. By the 2000s, breakthroughs in machine learning [LeCun et al. 2015] reignited interest in AI, bringing renewed attention to the importance of human factors. This resurgence, coupled with growing awareness of AI's societal impacts, underscored the need for systems that enhance human capabilities while upholding ethical considerations [Shneiderman

2020 2022]. Organizations such as the Partnership on AI [Partnership on AI 2016], the AI Now Institute [AI Now Institute 2017], and Stanford's Human-Centered AI Institute [Stanford's HAI 2019] have since expanded research into inclusive and accountable HCAI practices. The concept of HCAI has been explored from multiple perspectives, each emphasizing different aspects of its importance and application. For instance, [Shneiderman 2022] frames HCAI as both a process—focused on iterative refinement of user experience—and a product approach that enhances human performance while preserving user control. [Geyer et al. 2023] highlights the importance of transparency, equity, and ethical application, while [Interaction Design Foundation 2024] stresses the alignment of AI systems with human needs and values, accounting for social and cultural factors. Similarly, Stanford's HAI emphasizes the development of ethical AI that promotes human well-being, and [Vorvoreanu 2023] underscores the societal impact of AI, user agency, and responsible development. While these perspectives collectively advocate for supporting humanity, maintaining dignity, and prioritizing a human-centered approach, a comprehensive framework that systematically integrates and builds on these principles remains underdeveloped.

Human-centeredness lies at the heart of HCAI, emphasizing the design and development of AI systems that align with human values, needs, and ethical considerations. According to [Interaction Design Foundation 2024], human-centeredness involves designing systems that address the needs, preferences, and values of individuals while considering social and cultural factors. Rooted in disciplines such as HCI and UxD, human-centeredness prioritizes enhancing human well-being, fostering trust, and preserving user agency. Despite its foundational role, there is no consensus on what constitutes human-centeredness, and the key principles for systematically integrating it into AI development remain unclear. To address this gap, it is critical to understand how AI practitioners—who play a central role in designing and deploying human-centered AI systems—perceive and implement these principles. Their insights are essential for advancing a clearer and more actionable understanding of human-centeredness within HCAI, ensuring that AI technologies align with human needs and values in practice.

This research seeks to address these gaps by pursuing two key objectives. First, it aims to synthesize existing literature to conceptualize human-centeredness in HCAI and establish a clearer definition. Second, it explores and analyzes AI practitioners' perspectives on understanding and implementing human-centeredness in HCAI. By bridging theory and practice, this study aims to develop a unified framework that prioritizes user needs, upholds human values, fosters ethical principles, and preserves human autonomy, ensuring that AI advancements align with and support overall human well-being. The primary contribution of this work is the development of a hierarchical framework that systematically organizes 26 key attributes of human-centeredness into four tiers, validated through empirical data from AI practitioners. Unlike existing frameworks, which often focus on isolated aspects of HCAI (e.g., ethics, usability, or emotional intelligence), our framework integrates these dimensions into a cohesive structure that prioritizes ethical foundations, usability, emotional intelligence, and stakeholder engagement. This framework is the first to be empirically validated through practitioner input, offering actionable guidance for AI developers and researchers. By providing a clear, hierarchical structure, our work advances the field of HCAI by addressing the lack of consensus on what constitutes human-centeredness and offering a practical tool for designing AI systems that align with human values and needs. The research is guided by the following questions:

- How is human-centeredness conceptualized in HCAI based on insights from existing literature?
- What are the perspectives of AI practitioners on understanding and implementing human-centeredness in HCAI?

## 2 METHOD

This study employed a systematic, multi-step methodology to identify and validate attributes of "human-centeredness" in Human-Centered AI (HCAI). The process began with a comprehensive review of 81 definitions of HCAI sourced from academic papers, journals, international organizations, enterprise websites, educational videos, and other credible platforms. The sources were selected based on their relevance to HCAI and their contribution to understanding human-centeredness, with a focus on publications from the last decade (2013–2023) to ensure contemporary relevance. Using thematic analysis, recurring themes such as "human-centeredness," "responsible AI," and "ethical considerations" were identified. From this analysis, 78 unique attributes of "human-centeredness" were extracted and reviewed for relevance to the research context.

To further analyze the 78 attributes, a frequency analysis was conducted to measure how often each attribute appeared across the definitions. Attributes were categorized into three groups based on their occurrence: high-frequency, mid-frequency, and low-frequency. High-frequency attributes were those that appeared in more than 50% of the definitions, mid-frequency attributes appeared in 20–50%, and low-frequency attributes appeared in less than 20%. Low-frequency attributes, which were rarely mentioned, were excluded from further analysis. This process resulted in the identification of 15 high-frequency attributes and 63 mid-frequency attributes. The mid-frequency attributes underwent further evaluation by a panel of 15 domain experts, including AI researchers, HCI practitioners, and industry professionals. These experts rated each attribute on a 7-point Likert scale, ranging from "not at all important" (1) to "essential" (7). The average score of all mid-frequency attributes was calculated, and attributes were divided into two groups: those scoring above the mean and those scoring below. Attributes scoring above the mean were further refined through iterative consultations with the experts, resulting in the selection of 11 additional attributes deemed highly relevant to the study. These 11 attributes were combined with the 15 high-frequency attributes to form a final list of 26 attributes.

Each of the 26 attributes was carefully defined based on insights from the original 81 definitions, a supplementary literature review, and iterative discussions with experts. This process ensured that the final attributes were both conceptually robust and practically relevant. To validate the final list, a survey was conducted with AI professionals and domain experts. The survey was distributed through email, social media platforms (e.g., LinkedIn, Twitter), and instant messaging tools (e.g., WhatsApp, Slack). Participants rated the importance of each attribute on a 7-point Likert scale and provided qualitative insights through open-ended questions on defining HCAI, its role, and the challenges associated with achieving it. A total of 120 participants completed the survey, representing diverse geographic regions and professional backgrounds. Responses were analyzed to ensure that the final set of attributes accurately represented the key principles of human-centeredness in HCAI. Throughout the study, ethical protocols were followed to protect participant rights. This included obtaining informed consent from all participants, ensuring confidentiality of responses, and anonymizing data during analysis. The study was approved by the institutional review board (IRB) to ensure compliance with ethical research standards. This rigorous methodology culminated in a validated set of 26 attributes that highlight core principles of human-centeredness in HCAI, emphasizing user needs, human values, ethical considerations, and autonomy. Table 1 provides short definitions of each attribute.

Table 1: Short Definitions of the Attributes

**Human-centric design** prioritizes human values, needs, and experiences to create AI systems that enhance capabilities, uphold ethics, ensure transparency, and foster trust.
**User involvement in HCAI design** ensures their active participation, fostering intuitive, ethical solutions aligned with their needs and values.
**Human needs** encompass safety, usability, trust, and ethics, essential for AI systems to enhance well-being and daily life.

**User needs** are specific requirements and preferences essential for creating effective, intuitive AI systems aligned with their goals.
**Human values** encompass ethical principles like fairness, transparency, and privacy, ensuring AI systems uphold dignity and rights.
**Human experience** is the overall interaction and emotional response to AI systems, focusing on ease of use, satisfaction, and meeting user needs.
**Usability** is the ease of interaction with AI systems, ensuring they are intuitive, efficient, and effective in accomplishing tasks.
**Human factors** consider human capabilities, limitations, and behaviors to design AI systems that enhance performance, safety, and user satisfaction.
**Empathy** in human-centered AI involves understanding users' emotions and needs while creating systems that respond meaningfully and supportively.
**Human-centered design principles** prioritize user needs, usability, ethics, and well-being by integrating user feedback throughout development.
**User-centric data** focuses on consent, privacy, and relevance to enhance user experience and uphold ethical practices.
**Human well-being** ensures AI systems positively impact physical, mental, and emotional health, promoting safety and quality of life.
**Human trust** is users' confidence in AI systems' reliability, transparency, fairness, and ethics, ensuring security and respect.
**Human decision-making** supports informed, autonomous choices by providing clear, relevant, and unbiased information through AI systems.
**Feedback** involves gathering, analyzing, and using user input to iteratively improve AI systems, ensuring alignment with human needs, values, and expectations.
**Human control** ensures users retain authority to oversee, intervene, and direct AI system actions as needed.
**Stakeholder engagement** involves users, developers, and communities in AI design to address their needs and concerns.
**Human cognition** involves supporting users' mental processes like perception, memory, and decision-making to enhance human thinking with AI.
**Human behaviors** are user actions and patterns that AI systems should adapt to for intuitive and personalized interactions.
**User-friendliness** means designing AI systems that are intuitive, accessible, and easy to use for a positive experience.
**Human insights** are a deep understanding of user needs, preferences, and behaviors that guide AI development to enhance experiences.
**Human emotions** involve recognizing and responding to users' feelings for empathetic and effective AI interactions.
**Human goals** are user objectives that AI systems should support to achieve desired outcomes effectively..
**Human benefits** are the positive impacts AI systems provide, enhancing quality of life and well-being.
**User models** are computational representations of characteristics, preferences, and behaviors, used to personalize AI responses and interactions.
**Human dignity** ensures AI systems respect individual worth and rights, fostering fairness, privacy, and respect in interactions.

## 3 FINDINGS AND DISCUSSION

The participants in this study exhibit a diverse range of demographics and professional backgrounds. The average age of the participants is approximately 37.5 years, with a standard deviation of about 11.5 years, indicating a wide age distribution from early 20s to late 70s. Gender representation includes a majority of males and females, with a small percentage preferring not to disclose their gender. Participants hail from various countries, including Myanmar, Australia, Singapore, Germany, the UK, and the United States, reflecting a global representation. In terms of professional experience, participants have a broad spectrum of years in computing and AI-related fields, ranging from 0-1 years to over 16 years. Many are currently employed in corporate and industry roles, with positions such as Project Manager, Software Developer, Data Scientist, and Machine Learning Developer being common. Their knowledge of AI varies from basic to expert-level, with a significant number holding intermediate knowledge. This diverse group provides a comprehensive view of the industry, encompassing a wide range of experiences, expertise, and geographical backgrounds. A comprehensive analysis of AI practitioners' evaluations of 26 attributes defining "human-centeredness" in AI systems revealed a refined hierarchical structure of priorities. Using descriptive analysis of 7-point Likert scale responses, the 26 attributes were systematically categorized into four tiers based on mean scores: primary (highest priority), secondary (above-average importance), tertiary (moderate importance), and quaternary (lowest priority). This structured approach provides a clear framework for understanding the relative importance of each attribute, ensuring methodological rigor and offering valuable insights into how the field conceptualizes and practices human-centeredness in HCAI.

### 3.1 Primary Attributes: Ethical Foundations and Core Values in HCAI Development

The analysis reveals a clear prioritization of ethical and value-based attributes, with all primary attributes receiving high evaluations (M > 5.56 on a 7-point scale). These attributes underscore the centrality of ethics and human-centered values in HCAI development. 'Human values' emerged as the most critical attribute (M=5.91, SD=1.08), reflecting AI practitioners' commitment to embedding ethical principles such as fairness, transparency, and privacy in AI systems. This highlights the importance of aligning AI design with societal expectations for human dignity and individual rights, establishing ethics as the foundation of HCAI. 'Human benefits' (M=5.75, SD=1.07) received strong support, emphasizing the need for AI systems to deliver tangible positive impacts on individuals and society. This includes addressing fundamental needs such as safety, usability, trust, and ethics, ultimately enhancing quality of life, productivity, and societal progress. The low standard deviation indicates strong consensus on the importance of beneficial outcomes in HCAI.

'User-centric data' (M=5.73, SD=1.20) highlights the critical role of ethical data practices, including privacy protection, informed consent, and responsible data use. This attribute emphasizes the importance of ethical governance in ensuring data is managed with user consent and relevance, thereby enhancing user experience and trust. 'Human trust' (M=5.72, SD=1.15) emerged as a key attribute, underscoring the need for AI systems to be transparent, reliable, and aligned with user expectations. Trust is fostered through clear communication, predictable outcomes, and accountability, with the low standard deviation reflecting strong consensus on its essential role in HCAI. 'Human-centric design' (M=5.67, SD=1.20) emphasizes the prioritization of human values, needs, and experiences throughout AI development. This approach enhances human capabilities, promotes ethical practices, and ensures transparency, empowering users and fostering trust. 'Human needs' (M=5.66, SD=1.2) focuses on addressing fundamental requirements such as safety, well-being, and usability in AI systems. By meeting these needs, AI systems can deliver meaningful, user-focused outcomes that align with broader ethical considerations.

'Human-centered design principles' (M=5.63, SD=1.1) highlight the importance of an inclusive, iterative design process that addresses users' needs, limitations, and contexts. This approach embeds human perspectives throughout the AI lifecycle, ensuring solutions are relevant, accessible, and practical. Collectively, these primary attributes emphasize the centrality of ethics and value systems in HCAI. Attributes such as human values, trust, user-centric data, and human-centered design reflect a strong commitment to ethical principles, positive impacts, and fostering trust. By prioritizing human needs and values, AI systems can ensure transparent, accountable practices that enhance well-being and maintain trust throughout the AI lifecycle.

### 3.2 Secondary Attributes: Human-Centric Design and User Autonomy in AI Development

The secondary tier of attributes (5.38 ≤ M < 5.56) emphasizes human-centric design and operational aspects of HCAI, focusing on creating AI systems that are intuitive, aligned with user needs, and supportive of human autonomy. 'User-friendliness' (M=5.50, SD=1.24) highlights the need for intuitive and accessible AI systems, encompassing interface design, ease of interaction, and overall accessibility. This ensures positive and efficient user experiences for individuals with diverse technical backgrounds. 'Usability' (M=5.48, SD=1.11) emphasizes efficient and effective human-AI interactions, focusing on learnability, error prevention, and task completion efficiency. The moderate consensus reflects the importance of usability in successful AI implementation.

'User needs' (M=5.47, SD=1.02) underscores the importance of understanding and addressing user requirements and preferences. This ensures AI systems are effective, intuitive, and aligned with user goals, reflecting practitioners' commitment to meeting concrete expectations. 'Human decision-making' (M=5.45, SD=1.13) highlights the importance of supporting human autonomy by providing clear, relevant, and unbiased information, meaningful choices, and control over AI processes. This ensures AI systems support, rather than replace, human judgment, preserving human agency. 'Human dignity' (M=5.42, SD=1.30) bridges ethical principles and practical implementation, emphasizing respect for individuals' inherent worth and rights. It promotes fairness, privacy, and respect in AI interactions, though the higher standard deviation indicates diverse views on its application in practice. 'Human control' (M=5.38, SD=1.45) emphasizes user autonomy in AI interactions, ensuring users retain authority over AI systems through clear information, meaningful choices, and the ability to oversee or intervene as needed. The high standard deviation reflects varying practitioner views on the optimal level of oversight. Collectively, these secondary-tier attributes emphasize user-centered considerations. Attributes like user-friendliness, usability, and user needs focus on ease of use, efficiency, and alignment with expectations, while human decision-making, dignity, and control prioritize autonomy, informed choices, and individual rights. Together, they enhance human capabilities, uphold ethics, and ensure AI systems foster trust and transparency.

### 3.3 Tertiary Attributes: Emotional Intelligence and Human Factors in HCAI

The tertiary tier ($5.19 \leq M < 5.38$) addresses advanced aspects of human-centeredness in HCAI, focusing on refining AI systems to align with human needs, values, and experiences. Attributes such as human well-being, insights, user involvement, empathy, and human factors enhance AI systems' intuitiveness, supportiveness, and emotional awareness, fostering positive and engaging user experiences. 'Human well-being' (M=5.34, SD=1.42) emphasizes AI's role in enhancing users' physical, mental, and emotional health, promoting safety, happiness, and quality of life. While prioritizing well-being fosters security and satisfaction, the high standard deviation indicates differing views on its implementation.

'Human insights' (M=5.34, SD=1.11) highlights the use of a deep understanding of user needs, preferences, and behaviors to guide AI design, ensuring systems align with real-world expectations. The moderate consensus reflects recognition of the value of insights, though integration approaches vary. 'User involvement in the design process' (M=5.33, SD=1.20) underscores the importance of engaging users during AI development to create intuitive, ethical, and relevant systems. Incorporating user feedback improves satisfaction and usability, though the high standard deviation suggests diverse perspectives on the extent and methods of involvement. 'Empathy' (M=5.29, SD=1.40) focuses on understanding users' emotions, perspectives, and needs to create AI systems capable of recognizing and responding to human emotions. This fosters supportive interactions and stronger human-AI relationships, though the high standard deviation reflects diverse approaches to implementation.

'Human factors' (M=5.28, SD=1.14) emphasizes designing AI systems that consider human capabilities, limitations, and behaviors to enhance performance, safety, and satisfaction. By accommodating human strengths and constraints, practitioners ensure effective and harmonious technology use. 'Human experience' (M=5.25, SD=1.20) highlights the creation of positive, satisfying AI interactions by emphasizing ease of use and fulfilling user needs. Practitioners aim to meet functional requirements while leaving users confident and satisfied. 'Feedback' (M=5.25, SD=1.36) underscores the importance of gathering, analyzing, and integrating user input to iteratively improve AI systems. This ensures AI aligns with human needs, values, and expectations, though the high standard deviation reflects diverse approaches to collecting and applying feedback. Collectively, these tertiary-tier attributes refine human-AI interactions by aligning AI systems with human needs, values, and experiences. Attributes like human well-being, empathy, insights, and user involvement enhance intuitiveness and emotional awareness, fostering positive user experiences. Empathy and well-being emphasize AI's role in supporting users' health, while feedback and involvement drive continuous adaptation and relevance, demonstrating a commitment to creating responsive and emotionally considerate AI systems.

### 3.4 Quaternary Attributes: Behavioral and Emotional Intelligence in HCAI

The fourth tier of attributes (M < 5.19) represents elements that, while scoring above the midpoint on the 7-point scale, received relatively lower prioritization in HCAI development. These attributes emphasize personalization, emotional understanding, cognitive support, and stakeholder engagement, which, though valuable, were ranked as less critical compared to other attributes in shaping AI systems. 'Human goals' (M=5.17, SD=1.24) highlights the role of AI in supporting users' aims and helping achieve desired outcomes. By aligning with human objectives, AI enhances productivity and success in various tasks. The moderate standard deviation reflects differing practitioner approaches to aligning AI with user goals.

'Human cognition' (M=5.08, SD=1.17) emphasizes supporting users' mental processes, such as perception, memory, and decision-making. AI systems designed to complement cognitive abilities improve problem-solving and decision-making efficiency. The moderate standard deviation indicates diverse perspectives on integrating cognitive support into AI systems. 'Stakeholder engagement' (M=5.07, SD=1.31) stresses involving users, developers, and communities in AI design to address needs, values, and concerns. Collaboration fosters equitable, ethical, and user-centered systems, though the high standard deviation suggests varied approaches to effective engagement. 'Human behaviors' (M=4.98, SD=1.34) focuses on observing and adapting to user actions to create intuitive, personalized AI interactions. By dynamically understanding behavior, AI can tailor responses to meet individual needs, enhancing relevance and user experience.

'User models' (M=4.93, SD=1.13) involve computational representations of user characteristics, preferences, and behaviors, enabling AI to adapt dynamically for personalized interactions. This enhances relevance and effectiveness, providing customized support. 'Human emotions' (M=4.82, SD=1.48) focuses on recognizing and responding to users' feelings, fostering empathetic and meaningful AI interactions. By addressing emotional states, AI enhances user experiences with supportive, contextually appropriate responses. The high standard deviation reflects diverse views on integrating emotional intelligence into AI, highlighting different approaches and challenges in designing emotionally-aware systems. Collectively, these quaternary-tier attributes emphasize personalization, cognitive support, emotional understanding, and stakeholder involvement. Attributes such as human goals and cognition enhance productivity and decision-making by aligning AI with users' objectives and mental processes. Stakeholder engagement promotes equity and user-centeredness by incorporating diverse perspectives. Human behaviors and user models enable dynamic adaptation to individual preferences, while human emotions emphasize empathetic AI that fosters positive experiences. Together, these attributes contribute to adaptable, empathetic AI systems that support user aspirations and deliver effective, human-centered solutions.

### 3.5 Discussion

This study provides a comprehensive framework for understanding and prioritizing human-centeredness in HCAI by systematically categorizing 26 attributes into four hierarchical tiers. This structured analysis bridges theory and practice, emphasizing the integration of ethical, social, and human-centric values throughout the AI lifecycle. The primary contribution of this work is the development of a hierarchical framework that organizes 26 key attributes into four tiers—ethical foundations, usability, emotional intelligence, and stakeholder engagement—validated through empirical data from AI practitioners. Unlike existing frameworks, which often focus on isolated aspects of HCAI, our framework integrates these dimensions into a cohesive structure, offering actionable guidance for AI developers and researchers. By addressing the lack of consensus on human-centeredness, this work advances the field and provides a practical tool for designing AI systems that align with human values and needs.

The primary tier emphasizes ethical and value-driven attributes, such as human values, benefits, trust, and user-centric data, which received the highest ratings from practitioners. These findings highlight the necessity of embedding principles like fairness, transparency, and privacy into AI systems, aligning with [Shneiderman 2020]'s emphasis on ethical obligations in fostering societal trust. The secondary tier focuses on usability, user autonomy, and human control, reflecting the importance of designing intuitive and efficient AI systems that respect individual rights and promote informed choices. However, variability in practitioner responses, particularly for attributes like human control, suggests ongoing challenges in balancing user autonomy with system optimization.

The tertiary tier addresses emotional and cognitive dimensions, including well-being, empathy, and user involvement. While these attributes are recognized as valuable, their lower prioritization and greater variability in practitioner evaluations reflect the complexities of implementation. For instance, empathy and user involvement require nuanced approaches to capture diverse user experiences, highlighting the need for iterative and participatory design processes. The quaternary tier includes attributes such as human goals, cognition, emotions, and stakeholder engagement, which, while valuable, were ranked lower in terms of prioritization. These attributes highlight the potential of AI to support personalized, goal-oriented interactions, though their practical feasibility remains debated.

The hierarchical framework offers significant implications for both industry and research. For industry practitioners, it provides actionable guidance on prioritizing ethical foundations, usability, and emotional intelligence, enabling the design of AI systems that align with human values in domains such as healthcare, education, and autonomous vehicles. For policymakers, it underscores the need for regulations that promote ethical AI development while addressing practical challenges related to implementation and scalability. From a research perspective, this study advances the understanding of human-centeredness in HCAI by synthesizing insights from practitioners and existing literature into a unified framework. It highlights key areas for future research, including benchmarking human-centeredness, exploring cross-domain variations, and investigating longitudinal changes in practitioner priorities.

Despite its contributions, the study has limitations, including the underrepresentation of end users and policymakers, the influence of domain-specific contexts, and reliance on self-reported data. Future research should address these limitations by expanding participant diversity, conducting longitudinal studies, and triangulating findings with observational and experimental methods. In conclusion, this study's hierarchical framework provides a robust foundation for advancing human-centeredness in HCAI. By prioritizing ethical principles, human values, and trust, the framework aligns AI innovation with societal expectations while addressing practical challenges related to usability, autonomy, and emotional intelligence. We invite researchers and practitioners to apply this framework in diverse contexts and collaborate on refining its implementation, ensuring that AI systems continue to prioritize human well-being and align with societal values.

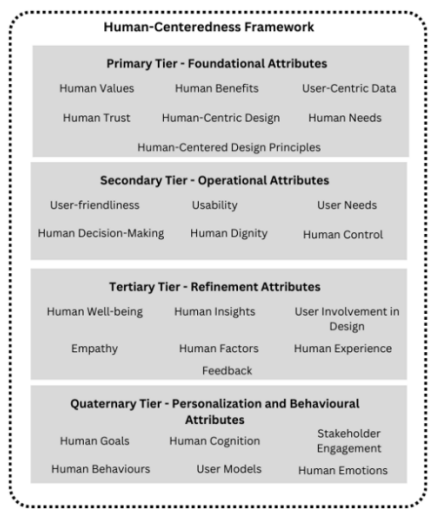

Figure 1. Human-Centeredness Framework